# Network Structure and Biased Variance Estimation in Respondent Driven Sampling*

**10-20-2015**

**Ashton M. Verdery[a]**
**Ted Mouw[b]**
**Shawn Bauldry[c]**
**Peter J. Mucha[d]**

* We thank Mason Porter for providing access to the Facebook data set we use. We also thank participants at the 2012 Hidden and Hard to Reach Populations Conference, Jonathan Daw, and Charles Seguin for helpful comments on earlier drafts. We are grateful to the Carolina Population Center for training (T32 HD007168) and general support (R24 HD050924) and to the Penn State Population Research Institute (R24HD041025). Peter J. Mucha was supported by the National Science Foundation (DMS-0645369) and by Award Number R21GM099493 from the National Institute of General Medical Sciences. This research uses data from Add Health, a program project directed by Kathleen Mullan Harris and designed by J. Richard Udry, Peter S. Bearman, and Kathleen Mullan Harris at the University of North Carolina at Chapel Hill, and funded by grant P01-HD31921 from the Eunice Kennedy Shriver National Institute of Child Health and Human Development, with cooperative funding from 23 other federal agencies and foundations. Special acknowledgment is due Ronald R. Rindfuss and Barbara Entwisle for assistance in the original design. Information on how to obtain the Add Health data files is available on the Add Health website (http://www.cpc.unc.edu/addhealth). No direct support was received from grant P01-HD31921 for this analysis. The content is solely the responsibility of the authors and does not necessarily represent the official views of any aforementioned funding agencies.

[a.] Corresponding author. Department of Sociology and Criminology, 716 Oswald Tower, The Pennsylvania State University, University Park, PA, 16802, USA. Email: amv5430@psu.edu

[b.] Department of Sociology, University of North Carolina at Chapel Hill, Chapel Hill, NC, USA.

[c.] Department of Sociology, University of Alabama at Birmingham, Birmingham, AL, USA.

[d.] Department of Applied Science, University of North Carolina at Chapel Hill, Chapel Hill, NC, USA.

**Network Structure and Biased Variance Estimation in Respondent Driven Sampling**

**Abstract**

This paper explores bias in the estimation of sampling variance in Respondent Driven Sampling (RDS). Prior methodological work on RDS has focused on its problematic assumptions and the biases and inefficiencies of its estimators of the population mean. Nonetheless, researchers have given only slight attention to the topic of estimating sampling variance in RDS, despite the importance of variance estimation for the construction of confidence intervals and hypothesis tests. In this paper, we show that the estimators of RDS sampling variance rely on a critical assumption that the network is First Order Markov (FOM) with respect to the dependent variable of interest. We demonstrate, through intuitive examples, mathematical generalizations, and computational experiments that current RDS variance estimators will always underestimate the population sampling variance of RDS in empirical networks that do not conform to the FOM assumption.  Analysis of 215 observed university and school networks from Facebook and Add Health indicates that the FOM assumption is violated in every empirical network we analyze, and that these violations lead to substantially biased RDS estimators of sampling variance.  We propose and test two alternative variance estimators that show some promise for reducing biases, but which also illustrate the limits of estimating sampling variance with only partial information on the underlying population social network.

**Network Structure and Biased Variance Estimation in RDS**

Respondent driven sampling (RDS) is a popular means of sampling difficult to survey populations. The ISI Web of Science database currently tags 642 academic articles with RDS listed as the topic [1]. These papers have been cited 10,217 times by 4,897 unique articles. A search of the NIH RePORTER database shows that the National Institutes of Health has awarded more than $180 million to 448 projects and subprojects with "respondent driven sampling" as a topic [2]. Much of this popularity owes to the fact that RDS is a cost effective and rapid means of sampling hard to reach populations, which have received increased attention across the social and health sciences.

There are two key components to the RDS approach. The first concerns sampling and recruitment, where respondents themselves are asked to find new survey participants through their social network connections with members of the target population, which are tracked with anonymous codes or coupons [3]. This is encouraged through a dual incentive structure where recruiters are paid for participating in the study and for recruiting others. The second component of RDS is inferential. Recruitment through social networks is complemented by a set of estimation techniques. Many of the estimation techniques used in RDS derive from the mathematics of random walks on graphs [4–6], because when RDS sampling and recruitment conforms to theoretical assumptions it mimics a simple random walk on an undirected, connected graph [7–9]. Under ideal conditions [10–12], RDS estimators of the population mean are asymptotically unbiased and generalizable to the population of interest, even absent a conventional sampling frame [7,13].

In this paper, we focus on an aspect of RDS inference that has received only limited attention in the literature to date: variance estimation. Most prior work on RDS inference focuses on estimating population means. Some have noted that RDS assumes sampling properties that are not followed in practice (e.g., non-branching recruitment, sampling with replacement, accuracy of degree reporting, an undirected network), which can lead to substantial biases [10,13–16]. Others have evaluated the precision of RDS mean estimates, or, more precisely, the variance in the sampling distribution of mean estimates ("sampling variance" [17]). An important recent finding is that RDS mean estimates may exhibit very high sampling variance compared to simple random sampling (SRS), even when assumptions are met [18]. This is an alarming finding for practitioners who typically collect only one sample, because their mean estimates may be far from the population mean, even if the average value from repeated sampling would converge to the population parameter.

Prior work has not thoroughly addressed the accuracy of RDS *estimates* of sampling variance, however. There are two commonly used estimators of sampling variance in RDS, known as the Salganik bootstrap estimator (SBE) [19], which uses a bootstrapping procedure to obtain variance estimates, and the Volz-Heckathorn estimator (VHE), which obtains variance estimates algebraically [7]. These approaches are quite similar, as both attempt to account for sample-induced correlations between cases that are close together in the referral network [14]. Such correlations lead the sampling variance of RDS to be larger than what would be obtained via SRS, yielding design effects greater than one, much in the same way that the design effects of cluster-based sampling increase as a function of intra-cluster correlations between units [20]. It is possible to obtain an exact variance estimator for random walks by incorporating data on the entire population's social network structure to account for these correlations [6]; we refer to this exact estimator as the Bassetti and Diaconis estimator. However, the RDS variance estimators lack data on the population network – they have only a sample– and as such need to approximate it. With a poor approximation, however, these variance estimators will be biased.

To date, despite attention to the general issue of sampling variance in RDS, the actual *estimators* of sampling variance used by researchers have escaped evaluation. The most thorough prior treatment  was by Neely [14], who diagnosed fundamental similarities between the SBE and VHE and limitations of both. Only two prior works have explicitly considered biased variance estimates in RDS [14,18]. Initial inquiries suggest that researchers should be wary when reporting confidence intervals and hypothesis tests based on commonly used RDS variance estimators. One study used simulated sampling and found that the SBE underestimates the empirical sampling variance using a set of school-based social networks [18], and two additional studies report that known population means are frequently much larger than the upper bounds of confidence intervals produced using the SBE [21,22]. However, the generality of these findings is uncertain. Whether they are limited to the cases studied or are endemic remains an open question.  In general, the sampling variance of random walks on graphs, and by extension other chain referral methods like RDS, depends on the specific network structure of the population under study which determines the closeness of nodes in the network, and, hence their covariance [6], and researchers do not know how closely the VHE or SBE can approximate it [14]. Because of this, it is important to assess why the RDS variance estimators might not reflect the population sampling variance, and whether this is a general problem or one that is primarily confined to specific types of "problem" networks.

In this paper, we diagnose when and why the current RDS variance estimators are biased and assess the generality of that bias across different types of networks. We build on Neely's [14] observation that both the SBE and the VHE rely, at a fundamental level, on a First Order Markov (FOM) assumption. This assumption holds that RDS recruitment can be modeled as a FOM process on the nodal attribute of interest, where transitions between states depend solely on the prior state and not a higher order sequence of prior states [23]. It is a convenient assumption for estimating RDS sampling variance, because a single RDS sample consists of a sequence of observed cases rather than the whole (population) network. Though the FOM assumption allows the VHE and SBE to estimate sampling variance from RDS sample data, researchers have not evaluated whether it is justifiable or assessed the potential consequences of its violation.

We organize the remainder of the paper as follows. We first provide a technical discussion of variance estimation in link tracing samples. We next extend this to RDS, wherein we explain the FOM assumption. We then offer two intuitive and general examples which demonstrate the VHE is biased when the FOM assumption is unwarranted. The first of these is a proof of concept and demonstrates the limitations the FOM assumption places on network structure. The second offers a simple model of correlated homophily which shows the VHE underestimates sampling variance when the FOM assumption is incorrectly applied, a problem which is likely to be quite general, at least whenever homophily based on an unaccounted for factor is present. After this, we test five variables in 215 empirical social networks for violations of the FOM assumption. We find that it is violated in every case for the full network data. However, when simulated samples are drawn from these data, these samples pass the test for FOM in the majority of cases. These results mean that, were a researcher in the field to have collected one of these simulated samples, she would not know, on the basis of the data collected, whether the RDS sampling variance estimators are likely to be biased. A fact which we confirm with our next set of analyses wherein we test whether sample-level FOM violations indicate especially poor estimates of the sampling variance (we find that they do not). This is a grave situation for RDS, whose mean estimators are known to exhibit high-sampling variance, because it indicates that researchers are unable to detect situations where mean estimates would have high variance and produce correspondingly large confidence intervals and other indicators of statistical uncertainty. Finally, we propose and test two modifications to the VHE that account for the branching structure of RDS data and higher-order Markov transition matrices and afford closer approximation of the actual RDS sampling process. These modifications improve sampling variance estimation in RDS, but they do not offer a complete solution.

This paper contributes to the literature on drawing inferences from network sampling designs by demonstrating that RDS sampling variance estimators are biased because their assumptions are invalid for many social networks and not just "unusual" or "difficult" ones. Our results also defy heuristic notions that situations where RDS estimates will exhibit high sampling variance can be easily detected on the basis of observed homophily, perceptions of network choke points, or sample level indicators of whether the network is FOM. They show that homophily is not strongly correlated with biases in variance estimates, that even though the networks we examine do not have choke points they still have high sampling variance, and that sample level indications that the network is not FOM give little indication of the extent of bias in estimates of sampling variance. Taken together, we show that RDS researchers using currently available estimators are unlikely to know – *a priori* or even after RDS data has been collected – whether a given network would exhibit high sampling variance. Such uncertainty calls for further development of procedures for accurately estimating RDS sampling variance.

### *1. Sampling Variance Estimation in RWS vs. SRS*

We can calculate nodal sampling probabilities at any sample size under a simple Markovian random walk model, which has the features of random referral, non-branching recruitment and with-replacement sampling assumed by RDS under ideal conditions [7,24]. These can be used to express the exact sampling variance of random walk sampling (RWS). In this section, we contrast how this calculation differs from analogous estimation in simple random sampling (SRS); we rely on a similar discussion in reference [14].

Let the matrix $G$ represent a population's underlying social network, which we assume comprises a single undirected connected component with at least one triangle. Ties between nodes $i$ and $j$ are indicated in cells $G_{ij}$ and $G_{ji}$ with a value of 1, while unlinked nodes have values of 0 in $G_{ij}$ and $G_{ji}$. A person's degree measures how many ties they have; the degree of person $i$ is given by $d_i = \sum_{j=1}^{N} G_{ij}$, where $N$ is the population size. If $D$ is a square diagonal matrix with $1/d_i$ along the diagonal, the Markov transition matrix is defined as $M = DG$, where $M_{ij} = G_{ij}/d_i$. If $P_0$ is a vector containing probabilities of starting the random walk in each node of the network, then $P_1 = P_0 M$ gives the probability of being in each node after one random walk step through the network, and $P_s = P_0 M^s$ gives probabilities of the random walk being at each node in the network after $s$ steps conditional on the starting probabilities [5].

It is convenient to express the transition matrix $M$ in terms of its eigenvalue decomposition. Equation (1) presents $M$ in terms of a set of eigenvalues and corresponding eigenvectors:

$$M = \sum_{k=1}^{N} \lambda_k D^{1/2} v_k v_k' D^{-1/2}, \qquad (1)$$

where $\lambda_k$ is the $k$th orthonormal eigenvalue of $M$ ordered in terms of magnitude $1 \ldots N$, $v_k$ is the $k$th eigenvector of a symmetrized version of $M$, and $D$ is the diagonal matrix of inverse degrees described above (primes denote the transpose operator)[1]. A key feature of the eigenvalue decomposition of $M$ is that the largest eigenvalue $\lambda_1 = 1$, and its corresponding eigenvector $v_1$ has elements $v_{1j} = \sqrt{\pi_j}$, where $\pi_j = d_j/2m$ is the steady state probability of being in node $i$, and $m = 1/2 \sum_i d_i$ is the total number of edges in the network [4]. After expressing $M$ in terms of the eigendecomposition in Equation (1), we can find the transition matrix after $s$ steps:

$$M^s = \left[ \sum_{k=1}^{N} \lambda_k D^{1/2} v_k v_k' D^{-1/2} \right]^s = \sum_{k=1}^{N} \lambda_k^s D^{1/2} v_k v_k' D^{-1/2}, \qquad (2)$$

where $v_k$ is the $k$th (right) eigenvector of $M^* = D^{-1/2} M D^{1/2}$ (see note 3). Using Equation (2), we see that $\lim_{s \to \infty} M^s = \lambda_1^s D^{1/2} v_1 v_1' D^{-1/2} = \pi_j$ because the first eigenvector has elements $v_{1i} = \sqrt{\pi_i} = \sqrt{d_i/2m}$ [5]. Importantly, $\pi_j$ is a probability vector where each element represents the probability of being in node $j$ in the steady state equilibrium. This is an important result because it means that the probability vector $P_s$ does not depend on $P_0$ when $s$ gets large. This is the source of the argument that the particular seeds from which an RDS sample begins will not bias estimates from large RDS samples, provided the sample size remains a small fraction of the population [10].

---

[1] The eigenvectors of $M$ will not necessarily be orthonormal because $M$ is not symmetric. However, as suggested by others [4], consider the matrix $M^* = D^{-1/2} M D^{1/2}$ where $D$ is the diagonal matrix of inverse degrees and $M$ is the Markov transition matrix defined above. This matrix is symmetric, as each cell $M^*_{ij} = \frac{G_{ij}}{d_i^{1/2} d_j^{1/2}} = M^*_{ji}$. $M^*$ can be decomposed as $M^* = VLV'$ where $L$ is a diagonal matrix consisting of the eigenvalues of $M^*$ along the diagonal and zeros off diagonal, and $V$ is a matrix of $M^*$'s eigenvectors. $M$ and $M^*$ share the same eigenvalues.

Sampling variance in RWS differs from the sampling variance of SRS in two important ways. The first owes to non-uniform sampling probabilities. The effects of this are most clearly understood by considering the probability model underlying SRS in terms of the matrix algebra introduced above. To get the sampling variance of SRS, we first arrange members of the population into a vector $Y$ with elements $Y = \{y_1, \dots, y_N\}$, and define $Y^*$ as its mean centered version $Y^* = \left\{ \left( y_1 - \frac{\sum_{i=1}^{N} y_i}{N} \right), (\dots), \left( y_N - \frac{\sum_{i=1}^{N} y_i}{N} \right) \right\}$. We denote the sampling variance of mean estimates of $Y$ from simple random samples as $\sigma_{\hat{\mu}}^2$, which can be estimated with the population variance and sample size as $\widehat{\sigma_{\hat{\mu}}^2} = \frac{1}{S} \sum_{i=1}^{N} \frac{Y^{*2}_i}{N}$, where $S$ indicates the sample size. If units are selected with given probabilities (in the vector $\pi$, where $\sum_i \pi_i = 1$) instead of with equal probabilities $(1/N)$, then the sampling variance is defined as:

$$\widehat{\sigma_{\hat{\mu}}^2} = \frac{1}{S}\sum_{i=1}^{N} \pi_i Y_i^{*2} = \frac{1}{S}\pi^{1/2}Y^*\left(Y^*\pi^{1/2}\right)'. \tag{3}$$

(Note that we present the matrix portion of Equation (3) in this way – breaking out the squared deviations into two pieces and taking the square roots of $\pi$ – to simplify algebra presented later).

The most important thing to note about Equation (3) is that it uses population values of $Y^*$ and $\pi$, despite being an estimate based on a sample. However, with a sample, researchers almost never know these values, which must be approximated. Denote the elements selected in a sample of size $S$ as $Y^{SRS} = \{y_1^{SRS}, \ldots, y_S^{SRS}\}$ and their sampling probabilities as $\pi_i^{SRS}$; assume $\sum_{i=1}^{S} \pi_i^{SRS} = 1$. Define the sample mean centered $Y$ values of the elements selected in the sample as $Y_i^{*SRS} = \left\{\left(y_1^{SRS} - \frac{\sum_{i=1}^{S} y_1^{SRS}}{S}\right), (\ldots), \left(y_S^{SRS} - \frac{\sum_{i=1}^{S} y_S^{SRS}}{S}\right)\right\}$. The estimated sampling variance on the basis of the sample is then:

$$\widehat{\sigma_{\hat{\mu}}^2} = \frac{1}{S-1}\sum_{i=1}^{S} \pi_i^{SRS}\left(Y_i^{*SRS}\right)^2 = \frac{1}{S-1}\pi^{SRS\,1/2}Y^{*SRS}\left(Y^{*SRS}\pi^{SRS\,1/2}\right)' \tag{4}$$

Equation (4) resembles the presentation in Equation (3), except that it now has sampled squared deviations $Y^{*SRS}$ (based on the sample estimate of the mean) and sampling probabilities $\pi^{SRS}$ (based on an estimate of these probabilities derived from the sample). Due to the Law of Large Numbers, Equation (4) provides an unbiased estimator of the sampling variance for SRSs where cases are selected with probability $\pi$ if $S$ is large; in other words $\widehat{\sigma_{\hat{\mu}}^2} \cong \sigma_{\hat{\mu}}^2$. In the case of SRS, sampling probabilities are uniform across units. In other sampling designs, however, $\pi$ can take on other values, and, indeed, in random walk samples $\pi_i^{RWS} = d_i/2m$ as per above, where $\pi_i^{RWS}$ indicates random walk sampling probabilities.

The estimation of sampling variance in link-tracing designs also differs from the analogous estimator in SRS because if there is homophily – the tendency for individuals with similar attributes to be friends with one another [25] – on the basis of some variable, then individuals in the network who are connected will tend to have similar values of that variable. As a result, values of the variable of interest from two cases in a random walk will be correlated if the number of steps between them is small. This results in non-zero covariances between cases in link-tracing style samples, which must be accounted for when estimating sampling variance. It is important to remember that homophily can exist on any sort of characteristic, observed (e.g., race) or unobserved (e.g., propensity to engage in risky behavior; [26]).

The variance of a RWS on a network can be analytically derived from the eigenvalue decomposition in Equations (2) and (3) [6]. First, working from Equation (3), we can express the population variance $\left((\pi^{SRS})^{1/2}Y^*\right)\left((\pi^{SRS})^{1/2}Y^*\right)'$ in terms of an orthonormal eigenvector basis $\{v_k : K = 1 \ldots N\}$, $\sum_{k=1}^{N} \alpha_k^2 v_k v_k' = \sum_{k=1}^{N} \alpha_k^2$, where $\alpha_k$ is a scalar constant that maps the $k$th eigenvector of the transition matrix $M^*$ onto $\pi^{1/2}Y^*$. Note that $v_k v_k' = 1$ and $v_j v_{k \neq j}' = 0$ because they are orthonormal. With this, we can denote the covariance between the $i$th and $j$th step of a random walk on $G$ as:

$$\text{cov}(i,j)_{\text{RWS}} = \sum_{k=2}^{K} \alpha_k^2 \lambda_k^{|j-i|}, \qquad (5)$$

where $\alpha_k$ is the mapping of the variable $Y^*$ onto the $k$th eigenvector of the transition matrix, $\lambda_k$ is the $k$th eigenvalue of the transition matrix, and $|j - i|$ is the number of steps between the $i$th and $j$th cases sampled by the random walk [6]. In general the covariance between two steps of a random walk is affected by all three components: $\alpha_k$, $\lambda_k$, and $|j - i|$. Using Equation (5), we can write an estimate of the sampling variance of a size $S$ random walk sample as the average of all the possible covariances among the population that the walk could take on $G$:

$$\widehat{\sigma^2_{\hat{\mu}RWS}} = \frac{1}{S^2(S-1)} \sum_{i=1}^{S} \sum_{j=1}^{S} \text{cov}(i,j)_{RWS}$$

$$= \frac{1}{S^2(S-1)} \sum_{i=1}^{S} \sum_{j=1}^{S} \sum_{k=2}^{K} \alpha_k^2 \lambda_k^{|j-i|}$$

$$= \widehat{\sigma^2_{\hat{\mu}}} + \frac{1}{S^2(S-1)} \sum_{i=1}^{S} \sum_{j=1, i \neq j}^{S} \sum_{k=2}^{K} \alpha_k^2 \lambda_k^{|j-i|}. \tag{6}$$

When $|j - i| = 0$, this reduces to the estimated simple random sampling variance of $Y$, $\widehat{\sigma^2_{\hat{\mu}}}$. Equation (6) highlights that the sampling variance of a RWS will be greater than the sampling variance of a same sized SRS if the variable $Y^*$ maps onto the eigenvectors such that the contribution in the sum from the positive eigenvalues outweighs that from the negative eigenvalues.

The other critical difference between Equations (6) and (4) is that the latter for SRS operates on a sample and estimates the sampling variance, while the former for RWS requires the underlying social network and dependent variable data from the entire target population. A researcher in the field, for example, would be unable to use Equation (6) without knowing the entire social network connecting individuals in the target population, which will obviously not be the case with a hidden population. In the next section, we discuss the VHE, which is an approximation of the Bassetti and Diaconis estimator used by the RDS literature to estimate the sampling variance of a respondent driven sample absent complete knowledge about the underlying social network.

## 2. Variance Estimation in RDS

In RDS the researcher does not have data on the network itself, only data on the sequence of recruitments that occurred and the characteristics of those recruited, i.e., a sample. We present the VHE in the same framework as equations (3)–(6) above. The key difference between the VHE and the exact estimator of Bassetti and Diaconis discussed above is that the estimator developed by Volz and Heckathorn uses the patterning of recruitments and characteristics in the sample in place of the entire population network. Given a RDS sample on the population connected by the social network $G$, let $Y^{RDS} \in Y$ indicate the cases sampled from the $Y$ values of the population. If $Y$ is dichotomous, let the matrix $C$ be a $2 \times 2$ matrix showing observed transition probabilities among values of $Y^{RDS}$ between respondents and those they referred to the study. If $Q_t$ is a $2 \times 1$ row vector indicating the probability that the $Y$ value of the $j$th respondent (or $j$th "step" in the interview) is 0 or 1, then we can estimate the Markov transition probability between any two steps $j$ and $i$ of the survey using observed categories of the dependent variable rather than nodes in the network as we did in Equation (6) above:

$$Q_s = Q_r C^{|j-i|}. \qquad (7)$$

We can then estimate the covariance between the $j$th and $i$th steps by modifying Equation (5) as:

$$\widehat{\text{cov}}(j,i)_{\text{RDS}} = \sum_{k=1}^{2} (\alpha_k^{VHE})^2 (\lambda_k^{VHE})^{|j-i|}, \qquad (8)$$

where the superscript VHE on alpha and lambda indicate that we are using $\alpha$ (the dependent variable $Y$ projected into the orthonormal basis) and $\lambda$ (the eigenvalues) from the eigendecomposition of sample observed category transition probabilities from the $2 \times$ 2 matrix $C$ instead of the population node transition probabilities from the $N \times N$ matrix $M^*$. This yields the VHE as[2]:

$$\widehat{\sigma^2_{\hat{\mu}RDS}} = \widehat{\sigma^2_{\hat{\mu}}} + \sum_{i=1}^{S}\sum_{j=1, i \neq j}^{S} \widehat{\text{cov}}(i,j)_{RDS}$$

$$= \frac{1}{S(S-1)}\sum_{i=1}^{S}\pi_i^{RDS}\left(Y_i^{*RDS}\right)^2 + \frac{1}{S^2(S-1)}\sum_{i=1}^{S}\sum_{j=1, i \neq j}^{S}\sum_{k=1}^{2}(\alpha_k^{VHE})^2(\lambda_k^{VHE})^{|i-j|}, \quad (9)$$

where $\pi_i^{RDS}$ is the RDS sampling probabilities.

Using $C$ instead of $M$ equates to making the FOM assumption because it assumes the likelihood of transitioning between categories of the $Y$ variable in question only depends on the category of the currently sampled node, one of several assumptions in RDS variance estimators described in detail in prior work [14]. Importantly, this is the

---

[2] Connecting Equation (9) to the notation used in Volz and Heckathorn [7], we can write

$\frac{1}{S(S-1)}\sum_{i=1}^{S}\pi_i^{RDS}\left(Y_i^{*RDS}\right)^2 + \frac{1}{S}\left(\frac{\sum_i^s Y_i^{RDS}/d_i^{RDS}}{\sum_i^s 1/d_i^{RDS}}\right)^2\left(1 - S + \frac{1}{\sum_i^s Y_i}\sum_{i=1}^{S}\sum_{j=1, i \neq j}^{S}C_{11}^{|j-i|}\right)$, where $Y_i^{*RDS}$ is the $Y$ value of the $i$th case selected mean-centered via the Volz-Heckathorn "RDS-2" mean estimator, $\pi_i^{RDS}$ is the corresponding case's sampling probability, $d_i^{RDS}$ is its degree, and $C_{11}^{|j-i|}$ is the estimated transition probability between $Y_{j,j \neq i} = 1|Y_i = 1$; this presentation can be found in reference [14]. Thus, $\frac{1}{S(S-1)}\sum_{i=1}^{S}\pi_i^{RDS}\left(Y_i^{*RDS}\right)^2$ is analogous to the estimated sampling variance of SRS (i.e., $\widehat{\sigma^2_{\hat{\mu}}}$ in equation 6), $\frac{\sum_i^s Y_i^{RDS}/d_i^{RDS}}{\sum_i^s 1/d_i^{RDS}}$ is the Volz-Heckathorn estimate of the sampling mean, and $\left(1 - S + \frac{1}{\sum_i^s Y_i}\sum_{i=1}^{S}\sum_{j=1, i \neq j}^{S}C_{11}^{|j-i|}\right)$ is the expected correlation between sampled units if the process is FOM.

same assumption made by the other commonly used RDS variance estimator, the SBE [14], which we evaluate via simulation later in the paper. The SBE is defined in the literature [14,19] as using a bootstrap sampling procedure to simulate synthetic RDS chains from the FOM approximation embedded in the $C$ matrix. Using $C$ to approximate the node-specific Markov transition probabilities is a simplification, as pointed out by its developers [7]. It may be a reasonable one to make because the $M$ matrix is unknown; however, the validity of this assumption has rarely been examined or tested in the RDS literature [14]. In the remainder of the paper, we explore the implications of the FOM assumption in greater detail.

### 3. What Happens To the VHE when FOM is Violated?

In this section, we provide several examples of what happens to RDS variance estimation via the VHE if the FOM assumption is violated.

#### Illustration 1: Intuition

We begin with an illustration that highlights how the relationship between sampling variance and sample size differs between FOM and non-FOM networks when using random walk sampling (RWS). This example is intentionally stylized so readers can see what is occurring and intuit the effects of network structure on the VHE of RDS sampling variance. Figure 1 shows two networks, A and B, where clear circles indicate nodes where $Y = 1$ and dark circles indicate nodes where $Y = 0$. These networks share several features: they have the same size, density, and degree distributions. In addition, Figure 1 was intentionally constructed so that both networks would have identical transition probabilities between categories of $Y$. This is important as it means that both networks have identical levels of dyadic homophily and that they share the $C$ transition matrix used by the VHE; that is, $C_A = \begin{bmatrix} 18/27 & 9/27 \\ 9/27 & 18/27 \end{bmatrix} = C_B$. Because $C$ is identical in these two networks, they will produce identical estimates of RDS sampling variance based on the VHE as shown in Equation (9).

-- Figure 1 about here --

However, there is one key difference between these networks: Network A is FOM with respect to $Y$ while Network B is not. This difference can be described in terms of the conditional probability of the current node's $Y$ value given the $Y$ values of prior nodes visited [23]:

$$\Pr(Y_s = 1 \mid Y_{s-1}, Y_{s-2}, \dots, Y_{s-\infty}) = \Pr(Y_s = 1 \mid Y_{s-1}). \tag{10}$$

Equation (10) holds for network A, and therefore network A is FOM, while for network B it does not hold (see below for a test of whether a network is FOM). Does this difference in network structure matter for RWS sampling variance? Given that the VHE estimates of sampling variance for these two networks will be identical (because they share the same $C$ matrix), if there is a large discrepancy between these networks in terms of the empirical RWS sampling variance (i.e., population sampling variance), then it suggests potential problems with the VHE.

Figure 2 shows how large an impact network structure can have on the sampling variance of random-walk based designs and that the VHE cannot detect these differences. Because we know the complete network structure, we can calculate the exact sampling variance of RWS estimates for these two networks using Equation (6). Alternatively, we can approximate it using the variance in the distribution of proportion estimates produced by repeatedly sampling the network with random walks starting from random points proportional to their stationary distribution probabilities a large number of times; these approaches yield indistinguishable results.

-- Figure 2 about here --

The striking result in Figure 2 is that while the sampling variance in Network A (the FOM network) approximates to the sampling variance of SRS as the sample size increases, the sampling variance in Network B (the non-FOM network) is much higher at every sample size. More importantly, while estimates from the VHE accurately describe the sampling variance of network A, they fail to do so for network B (the hollow circles on the graph show the VHE estimates run on *both* networks A and B). In fact, the VHE estimates for network B are identical to those obtained for network A, which makes sense as both networks have the same first-order transition probabilities. In terms of design effects (the ratio of RWS sampling variance to SRS sampling variance, where 1 indicates they are identical), the design effect from sampling $S = 5,000$ cases from Network A is 1.9997 while the same sample size in Network B yields a design effect of 29.2601 (the magnitude of this difference is roughly constant across sample sizes). In substantive terms, we can say that mean estimates based on RWS samples with 5,000 cases from Network B will be more almost 15 times more variable than the same estimates from Network A, yet, even with perfect information, the VHE will estimate that they have identical sampling variance. This finding is problematic for RDS because it indicates that if the FOM assumption is violated, researchers may have no idea whether their estimates of sampling variance – and, hence, their confidence intervals and hypothesis tests – are accurate or not.

*Illustration 2: Generality*

Though Figures 1 and 2 indicate that the VHE may produce estimates that are quite far from the population sampling variance of a network if the network is not FOM, one may wonder whether this result is produced by some feature of the two networks we considered. For instance, these are low degree networks and Network B has "choke points" (i.e., few paths between otherwise well connected clusters; see below for a more formal definition), which makes it the type of network that RDS heuristics suggest should be avoided (though, it is an open question whether a researcher or respondents themselves will know that the network of a hidden population has choke points in it and thus should not be surveyed with RDS). Because of these issues, we now turn to a more general illustration showing that the VHE will be downwardly biased – i.e., underestimate the population sampling variance – in all cases where there is homophily on an unobserved variable that is correlated with the $Y$ variable of interest.

Imagine a population composed of two groups categorized by a dichotomous $Y$ variable which are connected via a social network but which exhibit homophily on $Y$. We wish to take a RWS or RDS sample to estimate the population proportion of $Y = 1$. In addition, imagine there is a variable, $Z$, – unobserved by the researcher – which organizes a portion of the social ties in this network. As has been well documented, homophily exists on a wide range of dimensions, some of which may not be observed or anticipated by researchers or even research participants [25,26]. In this case, Z indicates a propensity for forming cross-group ties: individuals for whom $Z = 0$ only have ties with those within their $Y$ group, while those for whom $Z = 1$ have ties across $Y$ groups. For instance, if $Y$ indicated a dichotomous measure of race, then $Z$ might indicate an individual preference for interracial friends. Other examples that may generate a lack of conditional first-order independence are given on page 75 of reference [14].

For the sake of simplicity, assume an equal number of people of each $Y$ value in each of the $Z$ groups, and that the total degree of each $Y|Z$ combination is the same.[3] The number of friends among and between the different $Y|Z$ groups is depicted in Table 1. For example, there are $H$ friendships between people with different values of $Y$, while friendships among those with equal $Y$ values are marked in Table 1 as either $D$, $E$, or $F$, depending on the shared $Y$ value and the whether the ego and alter share $Z$ values and which $Z$ value they have. Because the total degree is the same for each $Y|Z$ combination we have $(D + E) = (E + F + H), \therefore D = (F + H)$.

-- Table 1 about here --

The transition matrix $M$ can be written as $M = \begin{bmatrix} 1-a & a & 0 & 0 \\ a & 1-a-b & b & 0 \\ 0 & b & 1-a-b & a \\ 0 & 0 & a & 1-b \end{bmatrix}$, where $a = \frac{F}{E+F+H}$ and $b = \frac{H}{E+F+H}$. $M$ represents a situation where there is heterogeneity in the level of dyadic homophily on $Y$. Individuals with $Z = 1$ form cross $Y$ friendships, while those with $Z = 0$ do not. Here $Z$ can represent anything that causes heterogeneity in mixing between $Y$s.

---

[3] Both of these assumptions can be relaxed without affecting the conclusions derived here. For example, we can allow for different numbers of people in each of the $Y|Z$ combinations and we can allow the degrees to vary. We have tested numerical examples under a variety of conditions (available upon request). The fundamental conclusion is the same, namely that homophily on an unobserved variable that is correlated with the dependent variable will lead to a biased VHE.

In contrast to the population transition matrix $M$, the VHE estimates sampling variance using the sample estimated $2 \times 2$ transition matrix $C$ as a function of the friendship propensities and the size of each group, $C = \begin{bmatrix} 1 - i & i \\ i & 1 - i \end{bmatrix}$, where $i = \frac{\text{cross } Y \text{ friends}}{\text{total friends}} = \frac{H}{2(E+F)} = \frac{b}{2}$. A critical result is that $C$ and $M$ have different eigenvalues. Because it is dimension two, $C$ has only two which are $\begin{bmatrix} 1 \\ 1 - b \end{bmatrix}$ when ordered by size. By contrast, $M$ has four eigenvalues which are $\begin{bmatrix} 1 \\ 1 - b - a + \sqrt{a^2 + b^2} \\ 1 - 2a \\ 1 - b - a - \sqrt{a^2 + b^2} \end{bmatrix}$ when ordered by size. Recalling Equations (5) and (6), for the VHE to accurately estimate the sampling variance of $Y$ via a random walk based approach like RDS, the second largest eigenvalues of these $C$ and $M$ would need to be equivalent. However, they are not, which means that when a correlated unobserved variable structures the homophily on $Y$, the VHE will not accurately estimate sampling variance. In fact, we can make a stronger claim in this case. Because $\left(1 - b - a + \sqrt{a^2 + b^2}\right) > (1 - b)$ for all nonzero $a$s and $b$s, we can say that the VHE will always underestimate the true sampling variance with this kind of network structure. This is a general result that compliments the intuition provided in Figures 1 and 2. There we showed that a random walk on a FOM network will mix more slowly than a random walk on a network with a higher order Markovian structure, but that the VHE will not be able to detect these differences. Slower mixing results in higher covariances between any two steps of a RWS or RDS sample drawn from that network, and, thus, higher sampling variance of mean estimates and larger design effects. However, the inability of the VHE to detect these differences – its biasedness – means that researchers will understate their uncertainty.

*Illustration 3: Computational Examples*

We now provide two concrete examples based on this illustration to demonstrate the effect of homophily on correlated unobserved variables.

For the first example, let $E = F = H = 10$, $M = \begin{bmatrix} .667 & .333 & 0 & 0 \\ .333 & .334 & .333 & 0 \\ 0 & .333 & .334 & .333 \\ 0 & 0 & .333 & .667 \end{bmatrix}$

and $C = \begin{bmatrix} .833 & .167 \\ .167 & .833 \end{bmatrix}$. In both transition matrices, the observed homophily between $Y$ groups is identical; that is, 16.7% of friendship ties are between those with $Y = 1$ and those with $Y = 0$. However, there is heterogeneity in the mixing between $Y$ groups defined by the unobserved $Z$ variable. Individuals in the first and fourth row of $M$ have no cross $Y$ mixing, while individuals in rows two and three have twice the average $Y$ mixing. In this example, the second largest eigenvectors are .804 and .667 for $M$ and $C$, respectively, indicating that a random walk on $M$ will reach equilibrium slower than a random walk on $C$. The standard deviation of a sample of 100 cases drawn from a random walk is 0.138 (design effect = 7.64) for $M$ and 0.110 (design effect = 4.88) for $C$. In other words, using the dyadic transition matrix $C$ in place of $M$ – i.e., using the VHE – results in a substantial underestimate of the true sampling variance.

As a second example, let $D = F = 100$ and $E = 10$, $M =$

$\begin{bmatrix} .952 & .048 & 0 & 0 \\ .048 & .476 & .476 & 0 \\ 0 & .476 & .476 & .048 \\ 0 & 0 & .048 & .952 \end{bmatrix}$ and $C = \begin{bmatrix} .762 & .238 \\ .238 & .762 \end{bmatrix}$. The second largest eigenvectors are .954 and .524 for $M$ and $C$, respectively. The standard deviation of a 100-case sample is 0.219 (design effect = 19.12) for $M$ and 0.0887 (design effect = 3.15) for $C$. In this second example, the observed homophily across $Y$ groups is lower than what was shown in the first example; here, 23.8% of friendships are cross group. However, $Z$ more fully structures the interaction of those within $Y$ groups – i.e., heterogeneity in cross $Y$ mixing – and this results in a dramatically higher difference in design effects for $M$ and $C$ (19.12 compared to 3.15, i.e., over 5 times larger).

*Summary of VHE Bias:*

The model of homophily on unobserved variables presented in this section is purposively simple in order to make analytical results tractable. Nonetheless, the basic intuition should be clear: if there is clustering within categories of the observed dependent variables—such as is evident in matrix $M$ of the second example above—then the VHE, which relies upon the observed transition matrix between categories of the variable of interest, $C$, will exhibit downward bias. The variance of a random walk is not just a function of dyadic homophily between different categories of the dependent variable, as both $M$ and $C$ have the same level of dyadic homophily but different design effects. In other words, it is network structure—not homophily on the observed, focal variable per se—that affects design effects and biases RDS sampling variance estimators downward [14]. Moreover, the examples presented here likely underestimate the role played by network structure as they focus on a simple set of networks and a limited $4 \times 4$ category transition matrix rather than a node level transition matrix that would be found in a real network. Indeed, the sampling variance of our computational examples could be correctly estimated by making a Second Order Markov assumption, but real world networks are unlikely to conform to that assumption as well. Below we test a modification to the VHE based on a second order assumption and show that while it frequently outperforms the classic VHE, it still does a poor job capturing the sampling variance of simulated RDS in empirical networks.

The fundamental point of this section was to show that if the FOM assumption is violated, as it is by the case of homophily on an unobserved variable, then the accuracy of the VHE derived estimates of RDS sampling variance are indeterminate, and will be downwardly biased under rather general conditions. As illustrated in the these examples, unbiased RDS variance estimators are predicated on the network being well described by homophily on a single observed category, and they are of little use when there is heterogeneity in the mixing among members of the groups defined by those categories. We now consider empirical data to evaluate the generality of these problems.

### 4. How often is FOM violated in empirical networks?

In this section, we ask whether researchers should generally expect networks to be well described by homophily on a single dimension, or, more specifically, how often they should expect the FOM assumption to be violated. The RDS literature has not explored this idea since its foundation [3]. We test the FOM assumption in 215 heterogeneous empirical networks drawn from two separate datasets. We then outline the issues faced by the two most used RDS variance estimators – the VHE and the SBE – when they are applied to networks that violate FOM, as we find that most empirical networks do.

*Data, Methods, and Measures*

Much prior methodological work in RDS has used simulated data [7,10,27]; however, it is challenging to accurately simulate all structural features found in empirical networks [28,29]. Because of this, we use data from the National Longitudinal Survey of Adolescent Health (Add Health) [30] and the Facebook 100 datasets [31–33]. These networks have been used in simulation based studies of network sampling performance [11,18]. In all networks, we restrict our analysis to individuals in the largest weakly connected components, and, in the Add Health data, we ignore the directionality of ties and treat all nominations as reciprocal. We use 115 networks from Add Health and 100 from the Facebook data set for a total of 215 empirical networks. In the Add Health networks, we test whether the FOM assumption holds for the following three dichotomous variables: race ($white = 1$), gender ($female = 1$), and sports participation ($yes = 1$). We look at the validity of the FOM assumption in two variables in the Facebook data: gender ($female = 1$), and class year ($freshmen = 1$).

These data are faithful to real world network patterns [34,35]. More importantly, they contain a diversity of network structures, which makes them excellent candidates for assessing the credibility of the FOM assumption in empirical networks and allows us to overcome criticisms that have plagued prior simulation work in RDS, namely that the empirical networks studied were too sparse, small, or contained "choke points". While these properties may characterize some of the Add Health networks, the Facebook networks we examine are not so limited. The best measure of choke points in a network is the average number of node independent paths. In any connected component, a set of nodes exists that, if removed, would disconnect that component. For a chain referral strategy to pass from one side of this nodal cutset to the other, it must pass through a node in this set. Menger's theorem [36] proves that the number of node independent paths in a graph is equal to the size of the smallest nodal cutset, which has been used to define the structural cohesion of a network [37]. We measured the numbers of node independent paths in the symmetrized largest connected components of the Add Health and Facebook networks used in this paper. The Facebook networks studied had an average number of node independent paths of 30.633 (with a range of 11.970 to 62.225), while the numbers in Add Health were smaller on average (mean of 4.884 with a range of 1.042 to 7.138)[4]. These macro-structural features, in addition to the high average degree ($FB = 51.640$, $AH = 6.971$), suggest that we study a substantial range of networks that are not limited by the heuristic notion of choke points[5]. The Facebook networks we examine are also quite large, orders of magnitude larger than the Add Health networks.

---

[4] Owing to the size of these networks and the computational complexity of calculating the number of node independent paths amongst all dyads in a network, we estimate the number of node independent paths in each network based on random samples of 10,000 dyads using maximum flow algorithms on the complete network. This provides asymptotically unbiased estimates.

[5] Other relevant statistics are as follows. The largest connected components in the Facebook networks ranged in size from 388-16,611 with a mean of 4,701. In Add Health, these numbers were 52-1,610 with a mean of 488. The proportions female in the Facebook networks ranged from 0.24-1.00 with a mean of 0.54, while the proportions freshmen in these networks ranged from 0.14-0.46 with a mean of 0.28. In Add Health,

The definition of a FOM process is as follows (see Equation 10 above):

$\Pr(Y_s = 1 \mid Y_{s-1}, Y_{s-2}, \ldots, Y_{s-\infty}) = \Pr(Y_s = 1 \mid Y_{s-1})$ [23]. Given this, a sufficient

condition that satisfies that the network is not FOM is:

$$\Pr(Y_s = 1 \mid Y_{s-1}, Y_{s-2}) \neq \Pr(Y_s = 1 \mid Y_{s-1}) \qquad (11).$$

By sufficient condition, we mean that if the preceding equation is true, then the network is not FOM. However, note that because this is only a sufficient condition, a failure to satisfy the preceding equation does not guarantee that the network is FOM. This makes it a conservative test: in cases where Equation 11 does not hold (i.e., the quantities are equal), the network may still not be FOM.

We test whether Equation 11 holds by estimating the following ordinary least

squares regression with robust standard errors for each of the variables of interest in each

of the networks from the Add Health and Facebook data sets:

$$\Pr(Y_s) = \beta_0 + \beta_1(Y_{s-1}) + \beta_2(Y_{s-2}) + \beta_3(Y_{s-1} \times Y_{s-2}) + \delta, \qquad (12)$$

where $\Pr(Y_s)$ is the proportion of an ego's alter's alters with $Y{=}1$, while $(Y_{s-2})$ and

$(Y_{s-1})$ are dichotomous indicators of whether ego's or alter's $Y{=}1$, respectively. Note that

we include ego himself as one of ego's alter's alters, which suffices to retain ego's alters

who lack alters (i.e., pendants) in the sample, and which makes sense for the with-

replacement process we study here. The resulting regression model thus contains one

observation for each edge in the network (or referral in the sample, depending on whether

---

the proportions female ranged from 0.01-0.69 with a mean of 0.54, while the proportions white ranged from 0.01-0.95 with a mean of 0.58, and the proportions participating in sports ranged from 0.28-0.95 with a mean of 0.56.

the analysis is conducted at the population or sample level, see below). Though each ego will have several alters in the data, and we make use of even more alters' alters in our definition of the dependent variable, the use of robust standard errors reduces concerns about clustering of the data. The sufficient condition shown in Equation 11 is true if the estimated coefficients for $\beta_2$ and $\beta_3$ are not jointly equal to 0, which we evaluate with the F-test of joint significance. Our null hypothesis is that the sufficient condition shown in Equation 11 is untrue – i.e., that $\Pr(Y_s = 1 \mid Y_{s-1}, Y_{s-2}) = \Pr(Y_s = 1 \mid Y_{s-1})$. While this does not guarantee that the networks are FOM, in cases where this test indicates we should reject the null hypothesis it means that the network is unlikely to be FOM because the current state depends on the prior state as well as how that state was reached.

  In addition to testing the FOM assumption in the complete networks, we also test it in RDS samples of size 200 on those networks conducted with replacement from a single seed drawn at random from the equilibrium distribution, because this is the type of data that a researcher who had collected a single sample might possess. We allow branching to occur where the probabilities of referring 0, 1, 2, or 3 new respondents in RDS are 1/3, 1/6, 1/6, and 1/3, respectively. Because this is the approach used in an influential past study [18], we focus on these results[6]. We conduct 500 simulated RDS samples in each of the 215 Add Health and Facebook schools, storing the relevant variables of interest.

---

[6]We also tested variants where we allow branching with the same probabilities as above, but the samples are conducted without replacement, and where we do not allow branching (both with and without replacement). We do not present these results but they did not alter our conclusions of substantial biases in the VHE and SBE.

After testing for FOM violations in these sampled network data, we then summarize some of the problems that the VHE and SBE estimators exhibit when applied to empirical network data. In each sample, we calculate the predicted proportion of $Y$ via the Volz and Heckathorn (i.e., the "RDS2 estimator") estimator of the mean $\hat{\mu}^{RDS} = \frac{\sum_{j}^{s} Y_j / d_j}{\sum_{j}^{s} 1 / d_j}$, where $d$ indicates degree [14,38]. We define the "population sampling variance" as the variance of the distribution of mean estimates obtained over $R = 500$ simulated samples in that network (which is approximately equal to what would be obtained via Equation 6 but is computationally feasible for larger networks); in other words,

$$\text{population sampling variance} = \frac{\sum_{r=1}^{R} \left( \hat{\mu}_r^{RDS} - \mu \right)^2}{R} = \sigma_{\hat{\mu}^{RDS}}^2, \tag{13}$$

where $r$ indexes the simulated replication of the sample (i.e., we simulate 500 replication samples in each empirical network). Defined in this way, the population sampling variance is the variance of the distribution of mean estimates across repeated samples. We use the population sampling variance to define the bias for the VHE and the other popular means of estimating sampling variance in RDS, the Salganik Bootstrap Estimator (SBE), which is

$$\text{bias} = \frac{\sum_{r=1}^{R} \left( \widehat{\sigma_{\hat{\mu}^{RDS}}^2} - \sigma_{\hat{\mu}^{RDS}}^2 \right)}{R} \tag{14}$$

where a value of zero indicates that the estimator is unbiased for that variable in that network[7]. The next quantity of interest is the ratio of the estimated sampling variance to the population sampling variance, which helps quantify how closely the VHE and SBE approximates the population sampling variance; thus, we also examine:

$$\text{ratio} = \frac{\sum_{r=1}^{R} \widehat{\sigma_{\hat{\beta}RDS}^2}/R}{\sigma_{\hat{\beta}RDS}^2} \tag{15}$$

in each network. Finally, though it may be biased, there is the possibility that the VHE estimates of sampling variance are highly correlated with the population sampling variance and thus researchers could simply inflate the variance estimator by some factor. To examine this possibility, we consider the correlation of the mean variance estimates in each network with the population sampling variance for each variable:

$$\text{correlation} = \text{cor}\left(\left(\sum_{r=1}^{R} \widehat{\sigma_{\hat{\beta}RDS}^2}/R\right), \sigma_{\hat{\beta}RDS}^2\right). \tag{16}$$

Finally, we note that the results about whether the network or sample is FOM pertain to whether or not researchers should expect that RDS estimators of sampling variance are underestimates (network level analysis) and can detect those cases (sample level analysis), as was demonstrated in the prior section. However, a different question is whether researchers can predict how large the underestimation bias in a given sample is likely to be rather than whether or not the estimators are biased toward underestimation.

---

[7] To gauge the potential influence of outlier RDS variance estimates on the relationship between the estimated sampling variance and the population sampling variance, we tested using the median estimate across the 500 simulations rather than the mean (not shown). This led to more severe biases and other problems than those reported in the manuscript.

Building from the literature reviewed in section 2, we know that the degree of bias is determined by higher order features of a network that is not FOM.

Echoing a general sentiment in the literature, it may be that homophily on the focal variable explains the degree of bias in cases where FOM is violated, so, to test this, we examine whether sample level homophily can predict levels of bias in cases where FOM is violated at the network level[8]. To do this, we compute the sample level homophily defined as the ratio of observed cross-group ties to expected cross-group ties in a given sample. We then regress bias on this measure of homophily to determine whether there is a meaningful and consistent relationship between bias and the homophily of a sample, which, if found, would indicate that the homophily observed in a sample can alert researchers to cases where bias is especially problematic. To facilitate interpretation, we focus on XY standardized regression models[9], where both the independent and dependent variables are standardized to have a mean of 0 and a standard deviation of 1. In XY standardized regression, the interpretation of coefficients is natural: a one standard deviation change in X (sample level homophily in the case we test) leads to a $\hat{\beta}$ standard deviation change in Y (sample level bias in the estimate of sampling variance in this case). To determine whether results owe to features of the networks or estimators we study or are general, we obtain parameters from regression models with and without absorbing indicators (i.e., fixed effects) for the network studied and for both the VHE and SBE estimators.

---

[8] A reviewer suggested this as a candidate explanation.

[9] We obtained substantively equivalent results in models run without XY standardization, but we focus on the XY standardized results because of their simpler interpretation in this case.

*Results*

The results of our tests of the FOM assumption on the complete networks are shown in panel A of Table 2, while results of the FOM test on individual samples are shown in panel B. Columns 1-3 show the proportion of FOM tests where we reject the null hypothesis that the network may be FOM under standard social science thresholds based on the F-test of joint significance ($p < 0.05$, $p < 0.01$, and $p < 0.001$). There are two key results. The first is that, for the complete networks (panel A), we reject the null hypothesis that the variable of interest in each network is FOM almost every single time. The one exception is a Facebook school where we could not calculate the FOM test for gender because the school is not co-educational. In other words, we find no cases where the fundamental premise of RDS variance estimation is a justifiable assumption. Worse still, the second key result in this table shows that, for the sample level tests (panel B), a near majority of the samples indicate the opposite, that the network may be FOM. This disjuncture indicates that it would be difficult for a researcher to know *a posteriori* whether the current methods of RDS variance estimation can be applied aptly. Though a given sample may seem appropriately characterized as FOM [3], the network from which it was drawn is highly unlikely to be. We return later to this disjuncture and its consequences for RDS variance estimation.

-- Insert Table 2 about here --

We next consider how problematic RDS variance estimation is when it is applied to empirical data whose complete network structure violates the FOM assumption. We look at the two most commonly used RDS variance estimators. Table 3 separate these estimators into two panels, with panel A showing the VHE and panel B showing the SBE. The first column shows the mean of the bias across the networks for each dataset and variable (Equation 14). The key point to notice about this column is that both the VHE and the SBE estimates are negatively biased in all cases. The second column shows the mean of the ratios of average VHE and SBE estimates of sampling variance in a given dataset and school to the population sampling variance (Equation 15). Most of the variables understate the true variance substantially – in the Add Health schools, the VHE estimated sampling variance understates the empirical sampling variance of RDS by about 85% – but this number ranges from as low as 55% to as high as 90% in the Facebook datasets. The SBE performs better here on average, with ratios of 0.2862 to 0.4421 in the Add Health schools and 0.1494 and 0.6544 in the Facebook ones. Finally, the third column of Table 3 shows the correlations (Equation 16), which highlight that there are substantial deviations from direct positive correlation and that the relationships between the average RDS estimates and the population values differ substantially by variable and dataset. We argue that this variation in correlations implies that researchers cannot know *a priori* whether the VHE or SBE estimates of sampling variance are useful.

-- Insert Table 3 about here --

The results in Table 3 are a conservative estimate of the problems with variance estimation in RDS. This is because they average, respectively, all VHE and SBE estimates across 500 RDS samples conducted in each school, which may paint an unrealistic picture of the practical utility of these estimators. Because researchers typically only collect one sample, we now turn to Figure 3 which shows box plots of the distribution across networks of the coverage rates by dataset, variable, and estimator (VHE vs. SBE). A given network's coverage rate is defined as the proportion of cases where the population mean $\mu$ is in the range $\hat{\mu}^{RDS} \pm 1.96 \sqrt{\widehat{\sigma^2_{\hat{\mu}^{RDS}}}}$, i.e., within the estimated 95% confidence interval. In SRS, the coverage rate is expected to be 0.95, but, as Figure 3 shows, the coverage rate for the VHE is substantially lower, and we see substantial variability in the distributions by data set and variable. For example, the "FB Freshman" variable shows that, on average, the 95% confidence interval for the VHE estimates of the mean proportion of freshmen contained the true mean in only 36.5% of the networks under study. This is substantial failure of confidence intervals for RDS. Beyond the poor coverage seen across all of the variables, a secondary point conveyed by Figure 3 is that the SBE generally outperforms the VHE.

<div align="center">-- Insert Figure 3 and Table 4 about here –</div>

With Table 4, we return to the disjuncture between population-level failure of the FOM test and sample-level passing of it that we noted in our discussion of Table 2. A natural question to ask is whether – *a posteriori* – a researcher can test her sample for FOM violations and discern whether the RDS variance estimators are likely to be biased. We split Table 4 into two panels: panel A shows the cases where the samples in Table 2 were found to not be FOM, while panel B shows those which may be FOM. The columns show the data set and variable combinations. Within each panel, we present the most relevant statistics averaged samples within that panel: the empirical design effect calculated from the population sampling variance in Equation 13 ("Mean empirical DE"), the VHE and SBE estimated design effects ("Mean VHE/SBE estimated DE"), and the coverage rates from both estimators ("Mean VHE/SBE 95% coverage rate"). The empirical design effects are generally smaller in the samples that may not be FOM; however, this is not true for the FB Freshman variable. However, the estimated DEs, using either the VHE or the SBE, do not appear appreciably closer to the population values (the empirical DEs). Neither do the coverage rates. Taken as a whole, these results suggest limited potential for sample-level FOM tests to be used as a diagnostic tool. Though researchers do not typically test for FOM violations, and while other, potentially yet-to-be-developed tests may be able to detect FOM violations in RDS sampled data, the most natural means of testing for FOM violations is unable to detect them. The development of methods to detect such violations thus represents a key area for potential research on RDS variance estimation.

Lastly, we estimated the parameters of XY standardized regression models for the relationship between sample level bias in the VHE and SBE estimators of RDS sampling variance and sample level homophily. These results are shown in Table 5. The key points highlighted in this table are that the relationship between sample level homophily and bias a) are in different directions across variables, b) are generally of low magnitude and often not distinguishable from 0 despite the large sample sizes, and c) differ between the VHE and SBE estimators. The conclusion to be drawn from these tests is that sample level homophily cannot be used to characterize the degree of bias in RDS estimators of sampling variance. These results show that another feature of networks that is commonly assumed to explain biases in RDS sampling variance estimators, level of homophily, is not a reliable indicator of whether the results of a single sample are biased.

-- Insert Table 5 about here --

In this section we found that the FOM assumption is routinely violated in empirical networks, but that researchers will not know this from the results of a single sample. Building on arguments introduced in Section 2, this finding indicates that the RDS variance estimators in common use can be expected to substantially underestimate the population sampling variance that RDS is likely to exhibit. We also examined a related question that helps contextualize the importance of our FOM results: how much do these violations of matter? More specifically, we explored the extent to which the current estimators for RDS sampling variance, the VHE and the SBE, are likely to underestimate the true sampling variance in these empirical networks. Our findings in this regard were surprising in two ways. First, Table 3 showed substantial downward biases in the VHE and SBE estimators of RDS sampling variance. It also showed little consistency across variables or data sets in the magnitude of this bias, or other properties of the relationship between the estimated values and the population parameter. This is important because it highlights that the current techniques of RDS variance estimation are wildly inaccurate, which makes sense because they are premised on a faulty assumption. In other words, this section has provided suggestive evidence that the core assumption underpinning variance estimation in RDS (the FOM assumption) is violated in a large proportion of empirical cases, that RDS variance estimators are biased in such circumstances, and that researchers will have difficulty knowing when this will be the case.

**Section 4. Improvements to the VHE.**

In this section, we test the performance of two improvements to the VHE. An easily diagnosable flaw in the VHE is that it fails to account for the branching nature of RDS data. As shown in Equations (5) and (6) above, the VHE uses the distance between sampled individuals $i$ and $j$, which in a random walk is equal to the number of steps between their appearances in the sample. However, in RDS, because of the branching structure of recruitment, these distance calculations will be more complicated. As such, the first estimator we introduce, based off of an earlier estimator developed by Neely in a prior investigation [14], which we call the "VHE with branching correction" (VHEwbc for short), explicitly tracks the network distance between individuals, so, if $i$ recruits $j$ who recruits both $k$ and $l$, we define the distance between $i$ and both $k$ and $l$ as 2. This approach should improve the VHE by more accurately calculating the covariance between cases.

The second improvement we test is relaxing the FOM assumption in the VHE. The VHE assumes the network is FOM with respect to the variable of interest because it uses a $2 \times 2$ transition matrix populated with the observed categorical transitions in the data. However, we can relax this assumption by making, e.g., a $4 \times 4$ transition matrix which is populated with the observed three step transitions (i.e., a second order chain); that is how often we see three-step chains with the following sequences of $Y$ values a) 0-0-0, b) 0-0-1, c) 0-1-0, d) 1-0-0, e) 0-1-1, f) 1-0-1, g) 1-1-0, or h) 1-1-1. This transition matrix encodes the probabilities by which the most recent pair of observed $Y$ values yield the next value; for example:

$$C_{2nd\ order} = \begin{bmatrix} \mathrm{Pr}_{000} & \mathrm{Pr}_{001} & 0 & 0 \\ 0 & 0 & \mathrm{Pr}_{010} & \mathrm{Pr}_{011} \\ \mathrm{Pr}_{100} & \mathrm{Pr}_{101} & 0 & 0 \\ 0 & 0 & \mathrm{Pr}_{110} & \mathrm{Pr}_{111} \end{bmatrix}, \tag{17}$$

where $\Pr_{000} = p(y_t = 0|y_{t-1} = 0|y_{t-2} = 0)$, that is the proportion of observed sequences of $Y$ values that went 0-0-0, and $\Pr_{001} = p(y_t = 0|y_{t-1} = 0|y_{t-2} = 1)$, and so on (note that 8 of the 16 elements in this transition matrix will be 0 by definition). In principle, one could further relax this assumption to incorporate even high-order chains,

but there is a tradeoff in terms of the number of such chains that one can observe in a

single sample. As such, we test whether incorporating higher order Markov assumptions

improves the validity of the VHE. In all cases, we also include the branching correction

(i.e., VHEwbc); we call this estimator the "VHE with higher order Markov", or VHEhom

for short. We test two higher order Markov assumptions: first we focus on a $2^{nd}$ order

assumption then we focus on a $3^{rd}$ order assumption. We also present results for the SBE

which adds another dimension of comparison with three variants of the algebraically based VHE estimator.

-- Insert Figure 4 about here --

Figure 4 presents the distribution of coverage rates across the different networks

for the VHE estimates of RDS sampling variance, for the VHEwbc estimates, for the

VHEhom estimates (note we did not calculate the VHEhom in the Facebook data set),

and for the SBE estimates. In most cases, the SBE outperforms the variants of the VHE

we tested in most cases. This is most clearly true for estimates of percent female in both

the Add Health and Facebook networks and the percent participating in sports in the Add

Health networks. The SBE's results are closer to the VHE variants for the race variable in the Add Health networks and worse than the VHEwbc estimator for the freshman variable in the Facebook networks. The proposed adjustments we consider generally improve the median coverage rate, but not substantially. Their effects also differ by variable and dataset. For the female variable in the Add Health networks, the VHEwbc improves estimates in all cases; the median, both quartiles and the outlier dots move closer to the desired 0.95. The VHEhom also improves estimates, if only marginally. For race in the Add Health networks, both the VHEwbc and the VHEhom outperform the VHE, but the VHEhom underperforms the VHEwbc. By contrast, for sports participation in the Add Health networks, the VHEhom substantially outperforms the VHEwbc. These cases illustrate that neither approach is significantly better than the other. In the Facebook 100 networks, we did not test the VHEhom owing to the size of these networks and the computational complexity of enumerating higher order chains, but suspect that the same general conclusions will hold. However, these networks are still interesting because they show just how much of a difference the VHEwbc can make. For the gender variable, there is almost no difference between the VHE and the VHEwbc. However, for percent freshmen, the difference is enormous with the $25^{th}$ percentile estimate from the VHEwbc higher than the $95^{th}$ percentile whisker from the VHE. On balance, however, the proposed corrections do not appear to substantially improve the coverage rates as none of median estimates are close to 0.95. On balance, researchers would be less likely to make inferential errors using the SBE estimator than any of the VHE variants we tested, but we note that they would still make the wrong inference frequently.

As a final illustration of the potential of these adjustments, we consider Figure 5. It plots differences between population design effects and the estimated design effects for the first (i.e., the VHEwbc), second (the VHEhom), and third order Markov strategies (not shown previously) for one variable (race) across the different networks in the Add Health data. The VHEwbc estimates are primarily found in the top left, and are often about an order of magnitude lower than the population DE. The VHEhom ($2^{nd}$ Order) is slightly closer to the line of parity, but still substantially different. The $3^{rd}$ Order Markov estimates are more scattered, but do not appear to be much better than the $2^{nd}$ Order estimates. Indeed, some are worse. This is because there is less and less data the higher order Markov process we estimate, and consequently additional error may be introduced by using higher order estimates. The reason for this fact is that there are fewer cases corresponding to each type of sequence the higher we go; as the cells become sparser, the precision with which they are estimated decreases.

-- Insert Figure 5 about here --

To summarize our analyses of potential corrections to the VHE, we note that the proposed corrections – more accurately accounting for the branching structure of the RDS chain rather than assuming a simple random walk and attempting to estimate higher order Markov transition patterns – do generally improve the variance estimation. However, the improvements we see are small and variable, and they do not improve coverage rates to a desirable level. Nonetheless, these procedures are a plausible first step toward improving estimates, and future work may improve on them. For instance, it may be that the eigensystem-based approach of the VHE fails with higher order Markov chains, but that a bootstrap approach more similar to the SBE would perform more desirably. We leave these questions for future work. More importantly, however, these results suggest the constraints that emerge from the typical RDS sampling methodology which focuses solely on recruiter-recruit links to the neglect of other relevant network data. We argue that a more fruitful approach may be to collect additional network data, either ego-networks [24,39–41] or more complete structures [11,42].

**Conclusions**

This paper has contributed to the literature on sampling hidden and hard to reach populations, and specifically Respondent Driven Sampling, by focusing on the issue of biased sampling variance estimation, which has only rarely been addressed to date [14,18]. Whereas prior work has documented biases in RDS mean estimators and the potential for RDS estimators to exhibit high sampling variance, the actual *estimation* of sampling variance has received considerably less attention. This is unfortunate for two reasons. First and most generally, if the RDS estimators of sampling variance are biased, then researchers cannot trust confidence intervals and hypothesis tests derived from these estimators. This is a problem for researchers and policy makers seeking to determine which populations have the highest disease prevalence, or whether observed changes in behaviors within a single population over time are due to actual changes or simply the variability that would be obtained through repeated sampling, to name two examples. Second, in the case of RDS, whose mean estimators are known to exhibit high sampling variance, an inaccurate means of estimating sampling variance will be especially problematic if it is downwardly biased. Our results suggest that the sampling variance estimators in use for RDS data are downwardly biased, indeed, massively so. Similar conclusions on a smaller scale have been highlighted in prior work [14,18]. We also found that the SBE generally outperforms the VHE or any natural extension of it, if only by a marginal amount.

Further, by focusing on the exact reasons for biases in the RDS variance estimators, this paper clarifies the heuristic notions prevalent in the RDS literature about which types of networks will be "problem cases" where RDS should not be applied. Unfortunately, our results demonstrate that such "problem cases" are common. Through mathematical illustrations, computational examples, and empirical analysis of 215 observed social networks from two different data sources, we have shown that the key assumption made by current RDS variance estimators – the First Order Markov assumption – is frequently violated. In addition, our results extend those of prior work [14,18] and show that the variance estimators perform poorly in many situations, and that the VHE as well as the SBE suffers this limitation. We examined two modifications to the VHE in an effort to reduce these biases, but, though they both offered some improvement, neither fundamentally solved the problem.

This paper has outlined new reasons that variance estimation in RDS needs more attention. Based on the performance of currently available estimators, a prudent researcher must wonder whether meaningful confidence intervals and hypothesis tests can be constructed. Given the results presented here, this does not appear to be the case because the variance estimators are so biased as to be effectively meaningless. However, further work may correct these issues, and other approaches to RDS estimation and diagnostic [10,13,24,39,40,43,44] and chain referral sampling [11,42,45–47] are being developed. These approaches, combined with renewed attention to the issue of estimating sampling variance in RDS, should pave the way for more accurate sampling variance estimation and a renewed emphasis on collecting additional network data as part of the sampling process.

**Figures and Tables**

**Figure 1. Networks with same degree distribution and proportion cross racial ties.**

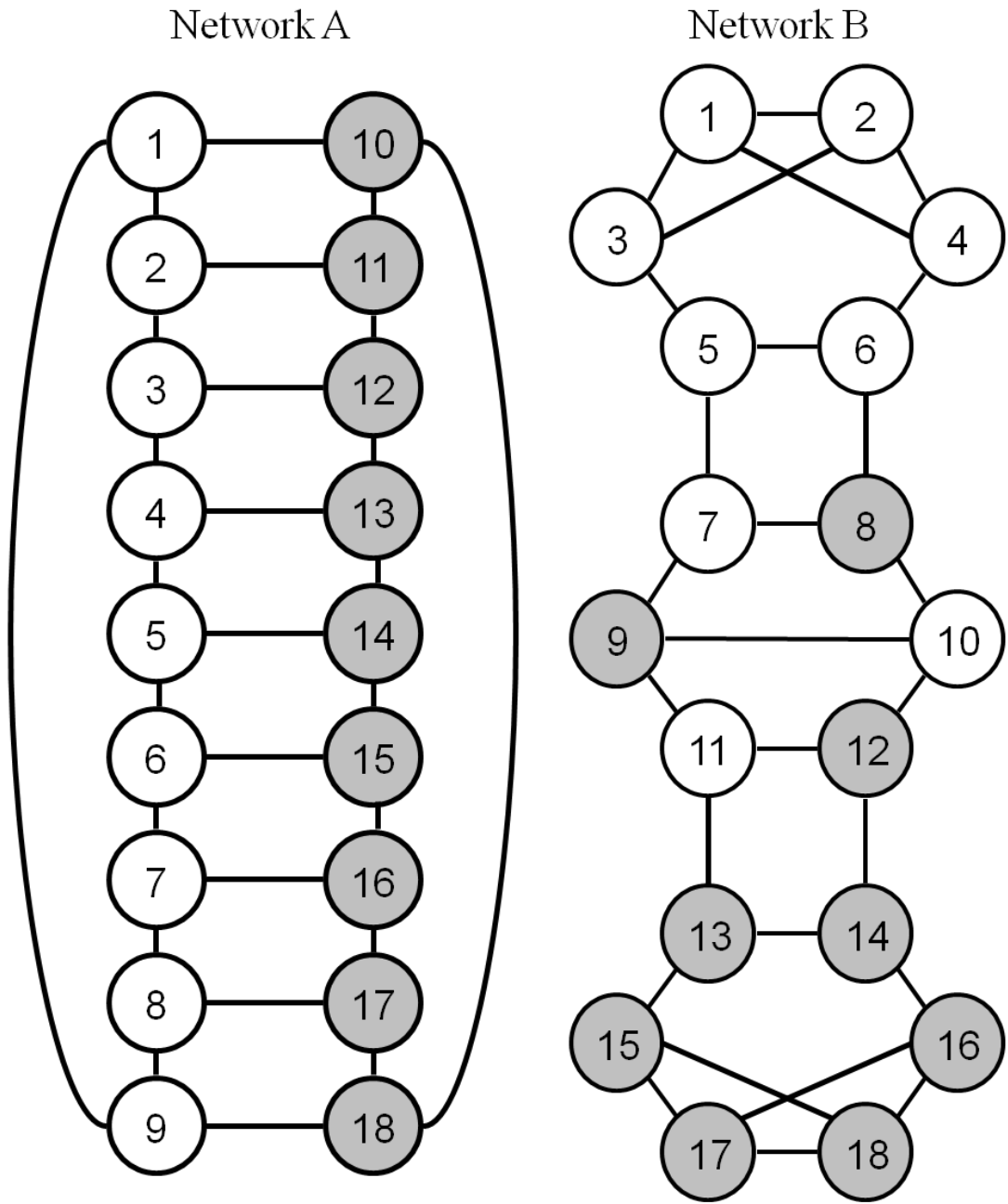

**Figure 2. Sampling variances on Networks A and B from figure 1, by method and sample size.**

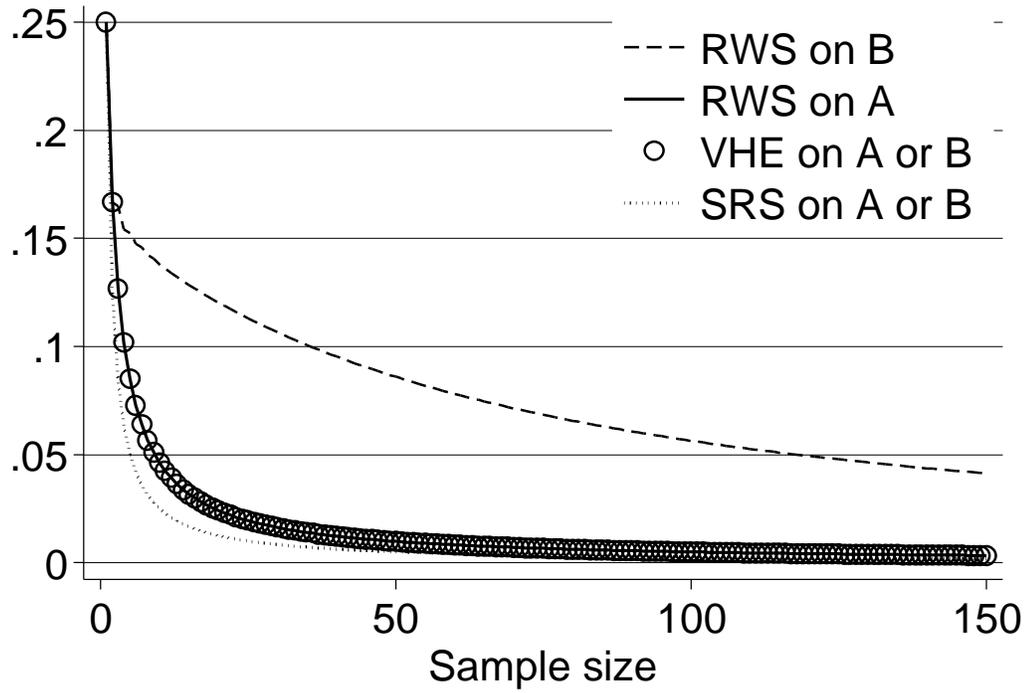

**Note: Like SRS, VHE produces identical estimates on networks A and B. For both networks, these estimates are identical to the population RWS sampling variance on A because network A is FOM.**

**Table 1. Algebraic representations of friendships between Y|Z groupings.**

| | | | Alter values | | | |
|---|---|---|---|---|---|---|
| | | | Y=0 Z=0 | Y=0 Z=1 | Y=1 Z=1 | Y=1 Z=0 |
| Ego Values | Y=0 | Z=0 | D | E | $0^1$ | $0^2$ |
| | Y=0 | Z=1 | E | F | H | $0^3$ |
| | Y=1 | Z=1 | $0^1$ | H | F | E |
| | Y=1 | Z=0 | $0^2$ | $0^3$ | E | D |

**Notes: Superscripts indicate no friendships in this cell because $^1$ $Z_{ego} = 0$ and $Y_{ego} \neq Y_{alter}$; $^2$ $Z_{ego} = 0 = Z_{alter}$ and $Y_{ego} \neq Y_{alter}$; $^3$ $Z_{alter} = 0$ and $Y_{ego} \neq Y_{alter}$.**

**Table 2. Descriptive statistics of First Order Markov (FOM) tests on Add Health and Facebook Networks, by analysis level and variable.**

| | (1) Pr. p<0.05 | (2) Pr. p<0.01 | (3) Pr. p<0.001 |
|---|---|---|---|
| *Panel A) Complete network FOM tests* | | | |
| Add Health Data Set | | | |
| Female | 1 | 1 | 1 |
| White | 1 | 1 | 1 |
| Sports | 1 | 1 | 1 |
| Facebook Data Set | | | |
| Female | 0.99 | 1 | 1 |
| Freshmen | 1 | 1 | 1 |
| | | | |
| *Panel B) Sample level FOM tests* | | | |
| Add Health Data Set | | | |
| Female | 0.567 | 0.367 | 0.182 |
| White | 0.556 | 0.409 | 0.245 |
| Sports | 0.640 | 0.450 | 0.248 |
| Facebook Data Set | | | |
| Female | 0.212 | 0.088 | 0.023 |
| Freshmen | 0.300 | 0.182 | 0.084 |

**Note: Pr. p<0.05, Pr. p<0.01, and Pr. p<0.001 indicate the proportion of networks in which the FOM test indicated we reject the null hypothesis that the network may be FOM. Note that in cases where the FOM test could not be calculated – e.g., non-coeducational schools or samples of only one gender – we considered this as indicating the network or sample may be FOM.**

**Table 3. Measures of the relationship between VHE and SBE estimates of sampling variance and the population sampling variance.**

| Data Set and Variable | (1) Bias | (2) Ratio | (3) Correlation |
|---|---|---|---|
| *Panel A) VHE Results* | | | |
| Add Health Female | -0.0092 | 0.1631 | 0.6375 |
| Add Health White | -0.0235 | 0.1498 | 0.9183 |
| Add Health Sports | -0.0111 | 0.1416 | 0.3333 |
| Facebook Freshman | -0.0496 | 0.0999 | 0.7851 |
| Facebook Gender | -0.0025 | 0.4981 | 0.4151 |
| | | | |
| *Panel B) SBE Results* | | | |
| Add Health Female | -0.0062 | 0.4421 | 0.6714 |
| Add Health White | -0.0215 | 0.2862 | 0.8380 |
| Add Health Sports | -0.0084 | 0.3868 | 0.4425 |
| Facebook Freshman | -0.0474 | 0.1494 | 0.7375 |
| Facebook Gender | -0.0012 | 0.6544 | 0.7380 |

**Note: Bias shows the mean of the average deviations between the sample estimates and the population parameters across replications and networks. Ratio shows the average ratio of estimated sampling variance to the population parameter. Correlation shows the correlation between the average of the sample estimates of sampling variance in each network and that network's population sampling variance.**

**Figure 3. Distribution across networks of coverage rates based on the VHE and SBE estimators, by variable and data set.**

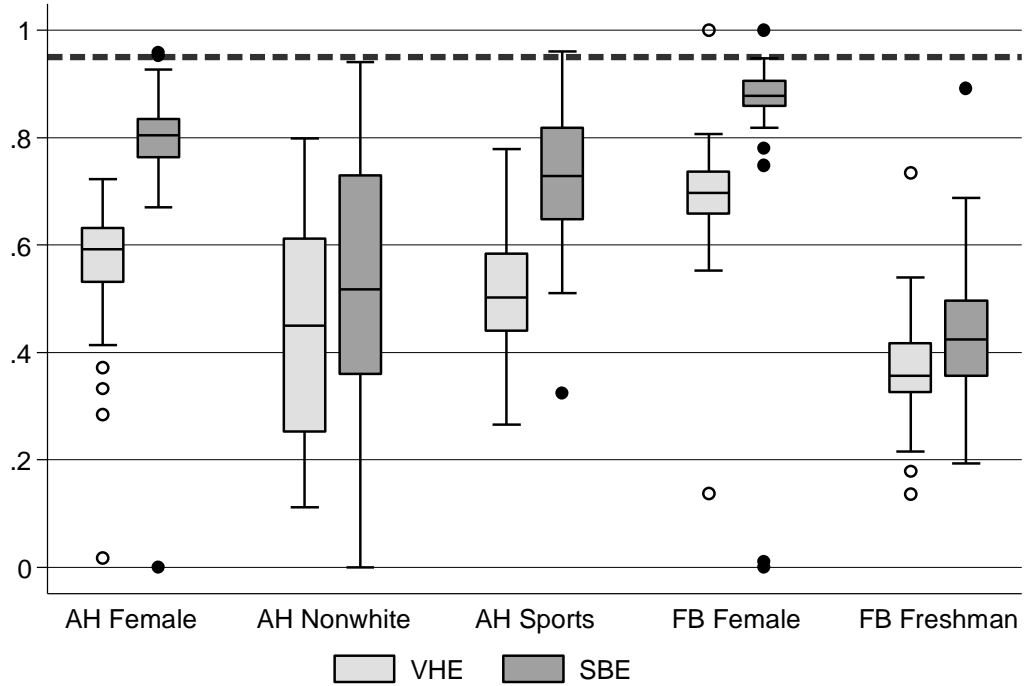

**Note: The expected coverage rate across networks for SRS is .95 (thick dashed line). AH indicates Add Health data set; FB indicates Facebook 100 data set. VHE indicates the Volz-Heckathorn Estimator, and SBE indicates the Salganik Bootstrap Estimator.**

**Table 4. Comparison of Design Effects, Estimated Design Effects, and Coverage Rates across samples that are not FOM or may be FOM, by data set and variable.**

| | Data Set and Variable | | | | |
| | AH Female | AH White | AH Sports | FB Female | FB Freshmen |
|---|---|---|---|---|---|
| *Panel A) Results on samples that are not FOM* | | | | | |
| Number of samples not FOM | 24,877 | 25,539 | 20,692 | 39,395 | 34,983 |
| Mean empirical DE | 7.906 | 24.697 | 9.921 | 2.856 | 60.260 |
| Mean VHE estimated DE | 1.302 | 3.590 | 1.816 | 1.450 | 11.035 |
| Mean SBE estimated DE | 3.568 | 4.364 | 3.488 | 1.821 | 10.454 |
| Mean VHE 95% coverage rate | 0.599 | 0.424 | 0.509 | 0.699 | 0.349 |
| Mean SBE 95% coverage rate | 0.803 | 0.426 | 0.715 | 0.863 | 0.398 |
| *Panel B) Results on samples that may be FOM* | | | | | |
| Number of samples may be FOM | 32,623 | 31,961 | 36,808 | 10,605 | 15,017 |
| Mean empirical DE | 9.386 | 28.185 | 11.441 | 3.045 | 51.756 |
| Mean VHE estimated DE | 1.392 | 1.908 | 1.323 | 0.769 | 4.836 |
| Mean SBE estimated DE | 4.322 | 4.869 | 3.935 | 1.929 | 8.568 |
| Mean VHE 95% coverage rate | 0.555 | 0.455 | 0.508 | 0.667 | 0.405 |
| Mean SBE 95% coverage rate | 0.785 | 0.588 | 0.734 | 0.861 | 0.505 |

**Notes: There are 57,500 samples for the 115 Add Health schools per variable; there are 50,000 samples for the 100 Facebook schools per variable. All statistics are averaged across samples in each panel.**

**Table 5. XY Standardized Regressions of Sample-Level Biases in Variance Estimates using VHE and SBE Estimators on Sample Level Homophily across the 5 Data Set/Variable Combinations Analyzed.**

| | VHE | VHE w/fixed network effects | SBE | SBE w/fixed network effects |
|---|---|---|---|---|
| Regression of variance estimate bias (standardized) for %: | | | | |
| 1) female variable in Add Health[a] | | | | |
| Sample homophily, standardized | 0.228*** | -0.095*** | 0.230*** | -0.006** |
| | [0.00] | [0.00] | [0.00] | [0.00] |
| R-squared | 0.052 | 0.998 | 0.053 | 0.776 |
| 2) non-white in Add Health[a] | | | | |
| Sample homophily, standardized | 0.448*** | -0.043*** | 0.473*** | 0.000 |
| | [0.00] | [0.00] | [0.00] | [0.00] |
| R-squared | 0.201 | 0.997 | 0.224 | 0.982 |
| 3) sports participants in Add Health[a] | | | | |
| Sample homophily, standardized | 0.143*** | -0.065*** | 0.218*** | -0.002 |
| | [0.00] | [0.00] | [0.00] | [0.00] |
| R-squared | 0.020 | 0.996 | 0.047 | 0.927 |
| 4) female in Facebook[b] | | | | |
| Sample homophily, standardized | -0.324*** | -0.229*** | -0.055*** | -0.002 |
| | [0.00] | [0.00] | [0.00] | [0.01] |
| R-squared | 0.105 | 0.760 | 0.003 | 0.332 |
| 5) freshmen in Facebook[b] | | | | |
| Sample homophily, standardized | -0.023*** | -0.136*** | 0.098*** | 0.001 |
| | [0.00] | [0.00] | [0.00] | [0.00] |
| R-squared | 0.001 | 0.936 | 0.010 | 0.902 |

**Notes: Standard errors in brackets. All regressions are based on XY standardized coefficients within estimator and data set/variable so all variables have a mean of 0 and a standard deviation of 1; models with fixed network effects mean dummy variables for each data set and network were absorbed by the model thereby removing network level differences. Constants not shown but all approximately 0 as would be expected in a XY standardized regression. \*\*\* p<0.001, \*\* p<0.01, \* p<0.05. [a] All models using Add Health data contain 57,500 simulated samples. [b] All models using Facebook data contain 49,000 simulated samples.**

**Figure 4. Distributions across networks for coverage rates based on the VHE, VHEwbc, VHEhom, and SBE estimators, by variable and data set.**

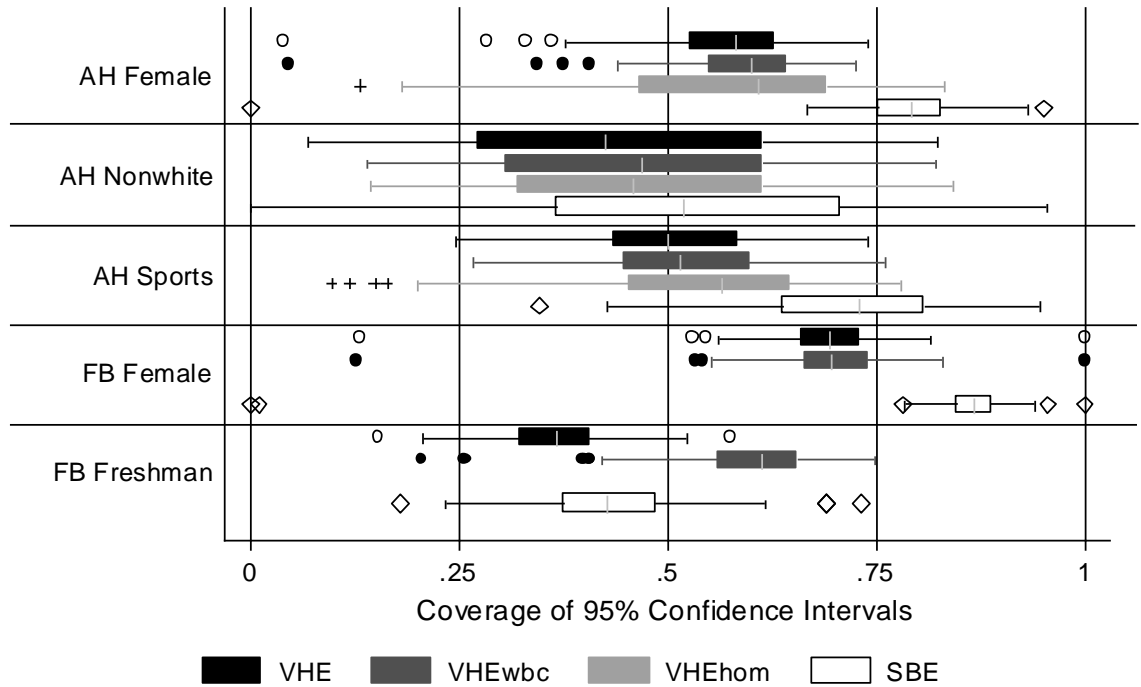

Coverage of 95% Confidence Intervals

VHE ■  VHEwbc ■  VHEhom ■  SBE □

**Note: The expected coverage rate across networks for SRS is .95. AH indicates Add Health data set; FB indicates Facebook 100 data set. VHE indicates the Volz-Heckathorn Estimator; VHEwbc indicates the VHE with Branching Correction; VHEhom is the VHE with Higher Order Markov assumptions; and SBE indicates the Salganik Heckathorn Estimator. The estimators are described in text.**

**Figure 5. Population RDS sampling variance vs. VHE estimated sampling variance with branching correction under different Markov Order assumptions, for Race in the AH data set.**

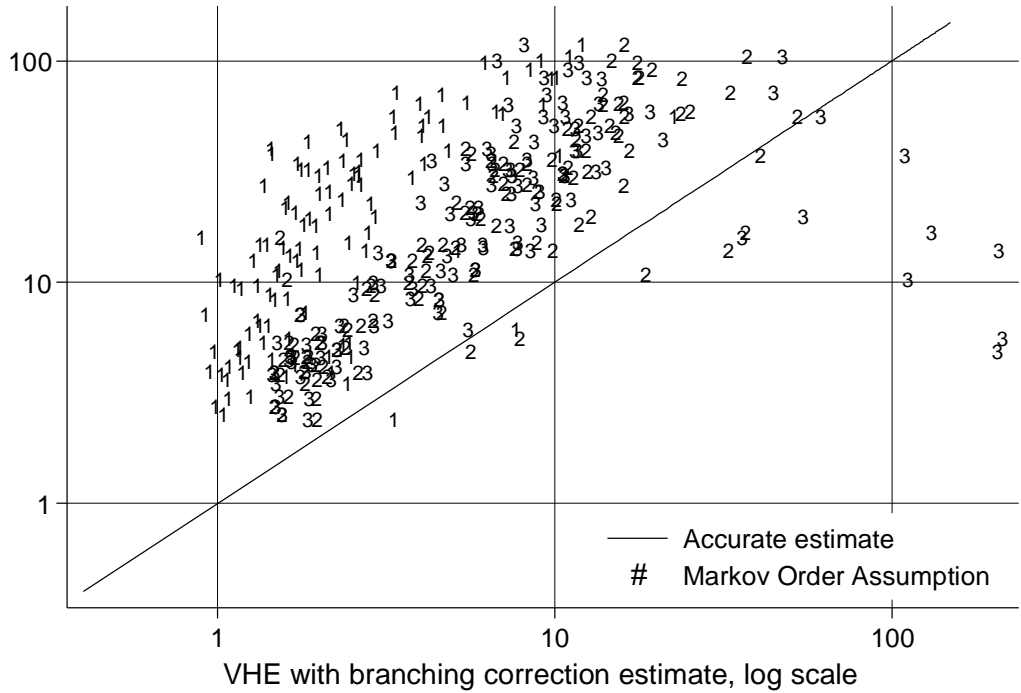

**Notes: Both scales in the figure are design effect scales and are logged. Symbols in the graph are as follows: 1 - FOM assumption (VHEwbc in figure 4); 2 - Second Order Markov assumption (VHEhom in figure 4); 3 - Third Order Markov Assumption (not shown in figure 4).**